\RequirePackage{lineno}
\documentclass[twocolumn,aps,prc,showpacs,superscriptaddress,floatfix]{revtex4-1}
\usepackage{rotating}
\usepackage{epsfig}
\usepackage{hyperref}
\usepackage{lineno}
\usepackage{url}
\usepackage{multirow}
\usepackage{float}

\usepackage{amsmath}
\usepackage{graphicx}
\usepackage[caption=false]{subfig}

\graphicspath{ {./}, {./plot/}, {./image/} }
\DeclareGraphicsExtensions{ .pdf, .png}

\makeatletter
\renewcommand{\p@subsection}{}
\renewcommand{\p@subsubsection}{}
\makeatother

\makeatletter
\let\LN@equation\equation
\let\LN@endequation\endequation
\renewcommand{\equation}{\linenomath\LN@equation}
\renewcommand{\endequation}{\LN@endequation\endlinenomath}
\let\LN@gather\gather
\let\LN@endgather\endgather
\renewcommand{\gather}{\linenomath\LN@gather}
\renewcommand{\endgather}{\LN@endgather\endlinenomath}
\makeatother

\begin{document}


\newcommand{\der}{\text{d}} 
\newcommand{\nlp}{n_{\text{W}}^{+}}
\newcommand{\nln}{n_{\text{W}}^{-}}
\newcommand{\nrp}{n_{\text{E}}^{+}}
\newcommand{\nrn}{n_{\text{E}}^{-}}
\newcommand{\Nlp}{N_{\text{W}}^{+}}
\newcommand{\Nln}{N_{\text{W}}^{-}}
\newcommand{\Nrp}{N_{\text{E}}^{+}}
\newcommand{\Nrn}{N_{\text{E}}^{-}}
\newcommand{\nos}{n_{\text{OS}}}
\newcommand{\nss}{n_{\text{SS}}}
\newcommand{\vvp}{v_{2,\pi^+}}
\newcommand{\vvn}{v_{2,\pi^-}}
\newcommand{\gos}{\gamma_{\text{OS}}}
\newcommand{\gss}{\gamma_{\text{SS}}}
\newcommand{\gdf}{\Delta\gamma}
\newcommand{\gdfbb}{\Delta\gamma_{\text{BB}}}
\newcommand{\rbb}{r_{\text{BB}}}
\newcommand{\pbb}{\varphi_{\text{BB}}}


\title{Back-to-back relative-excess observable in search for the chiral magnetic effect}

\author{Yicheng Feng}
\email{feng216@purdue.edu}
\address{Department of Physics and Astronomy, Purdue University, West Lafayette, IN 47907, USA}

\author{Jie Zhao}
\email{zhao656@purdue.edu}
\address{Department of Physics and Astronomy, Purdue University, West Lafayette, IN 47907, USA}

\author{Fuqiang Wang}
\email{fqwang@purdue.edu}
\address{Department of Physics and Astronomy, Purdue University, West Lafayette, IN 47907, USA}
\address{School of Science, Huzhou University, Huzhou, Zhejiang 313000, China}

\date{\today} 


\begin{abstract}

\begin{description}
	\item[Background]
	The chiral magnetic effect (CME) is extensively studied in heavy-ion collisions at RHIC and LHC.
	In the commonly used reaction plane (RP) dependent, 
	charge dependent azimuthal correlator ($\Delta\gamma$), 
	both the close and back-to-back pairs are included.
	Many backgrounds contribute to the close pairs (e.g. resonance decays, jet correlations),
	whereas the back-to-back pairs are relatively free of those backgrounds.
	\item[Purpose]
	In order to reduce those backgrounds,
	we propose a new observable which only focuses on the back-to-back pairs, 
	namely, the relative back-to-back opposite-sign (OS) over same-sign (SS) pair excess ($\rbb$)
	as a function of the pair azimuthal orientation with respect to the RP ($\pbb$).
	\item[Methods]
	We use analytical calculations and toy model simulations
	to demonstrate the sensitivity of $\rbb(\pbb)$ to the CME and its insensitivity to backgrounds.
	\item[Results] 
	With finite CME, 
	the $\pbb$ distribution of $\rbb$ shows a clear characteristic modulation.
	Its sensitivity to background is significantly reduced compared to the previous $\gdf$ observable.
	The simulation results are consistent with our analytical calculations.
	\item[Conclusions]
	Our studies demonstrate that the $\rbb(\pbb)$ observable is sensitive to the CME signal 
	and rather insensitive to the resonance backgrounds.
\end{description}

\end{abstract}



\maketitle


\section{Introduction}

In quantum chromodynamics (QCD), vacuum fluctuations can produce nontrivial topological gluon fields in local domains~\cite{Lee:1974ma}.
The chirality of quarks, under the approximate chiral symmetry, is imbalanced in those gluon fields~\cite{Morley:1983wr,Kharzeev:1998kz,Kharzeev:2007jp}.
This violates the $\mathcal{CP}$ symmetry in QCD in local domains.
In a strong magnetic field, the single-handed quarks will polarize along or opposite to the magnetic field depending on the quark charge. 
This produces an electric current along the magnetic field,
resulting in an observable charge separation in the final state~\cite{Kharzeev:1998kz,Kharzeev:2007jp}.
This phenomenon is called the chiral magnetic effect (CME)~\cite{Kharzeev:1998kz,Kharzeev:2007jp}.

In non-central heavy-ion collisions, the spectator protons can produce an intense, transient magnetic field, 
approximately perpendicular to the reaction plane 
(RP) (spanned by the beam direction and the impact parameter)~\cite{Kharzeev:2007jp}.
The high energy density region created in these collisions, 
where the approximate chiral symmetry may be restored, 
provides a suitable environment to search for the CME~\cite{Kharzeev:2007jp}.
The observation of CME-induced charge separation in heavy-ion collisons would provide a strong evidence
for QCD vacuum fluctuations and local $\mathcal{CP}$ violation.

The CME is extensively studied in heavy-ion experiments at the Relativistic Heavy Ion Collider (RHIC)~\cite{Abelev:2009ac, Abelev:2009ad, Adamczyk:2014mzf, Adamczyk:2013hsi, Zhao:2017wck,Zhao:2017ckp, Zhao:2019qm, Adamczyk:2013kcb} and the Large Hadron Collider (LHC)~\cite{Khachatryan:2016got, Abelev:2012pa, Acharya:2017fau, Sirunyan:2018ac}. 
To probe the CME signal, the RP-dependent, charge-dependent $\Delta\gamma$ observable was proposed~\cite{Voloshin:2004vk} and widely used.
Positive CME-like signals in $\gdf$ have been observed in both heavy ion collisions (Au+Au at RHIC~\cite{Abelev:2009ac,Abelev:2009ad,Adamczyk:2014mzf,Adamczyk:2013hsi} and Pb+Pb at the LHC~\cite{Abelev:2012pa}) and small systems collisions (p+Au and d+Au at RHIC~\cite{Zhao:2017wck,Zhao:2017ckp} and p+Pb at the LHC~\cite{Khachatryan:2016got}),
where the latter is believed to come only from backgrounds.
In fact, it has been pointed out previously that the $\gdf$ in heavy-ion collisions
was contaminated by major backgrounds~\cite{Wang:2009kd,Bzdak:2009fc,Schlichting:2010qia}.
Various methods have been developed to suppress the backgrounds, 
such as event shape engineering~\cite{Sirunyan:2018ac,Acharya:2017fau}, 
invariant mass dependence~\cite{Zhao:2017wck},
and the comparative $\gdf$ measurements with respect to the reaction and participant planes~\cite{Xu:2018prl, Xu:2018cpc}.
The current results with those methods show a CME signal consistent with zero.

In this paper, we propose a new method.
In the original definition of $\gdf$, both the close pairs and the back-to-back pairs are included.
Many backgrounds contribute to the close pairs (e.g. resonance decays, jet correlations)~\cite{Wang:2009kd, Bzdak:2009fc, Liao:2010nv, Bzdak:2010fd, Schlichting:2010qia, Pratt:2010zn, Petersen:2010di, Toneev:2012zx, Zhao:2018ixy, Zhao:2018skm, Zhao:2019hta},
whereas the back-to-back pairs are relatively free of those backgrounds.
Thus, we propose a new observable which only focuses on the back-to-back pairs, namely, the relative back-to-back opposite-sign (OS) over same-sign (SS) pair excess
as a function of the pair azimuthal orientation with respect to the RP. 
We use simulations by a toy model (previously used in Ref.~\cite{Wang:2016iov, Feng:2018so}) 
to demonstrate the sensitivity of this observable to the CME signal and insensitivity to the backgrounds.
The relationship between this new observable and the $\Delta\gamma$ observable is also discussed.

The paper is organized as follows.
Section~\ref{Methodology} describes the methodology of this study.
Section~\ref{Results} shows our toy-model simulation results
using the new method.
Section~\ref{Summary} summarizes this work.


\section{Methodology} \label{Methodology}

\subsection{New back-to-back relative-excess observable, $\rbb$}

\begin{figure}
	\includegraphics[width=1.0\linewidth]{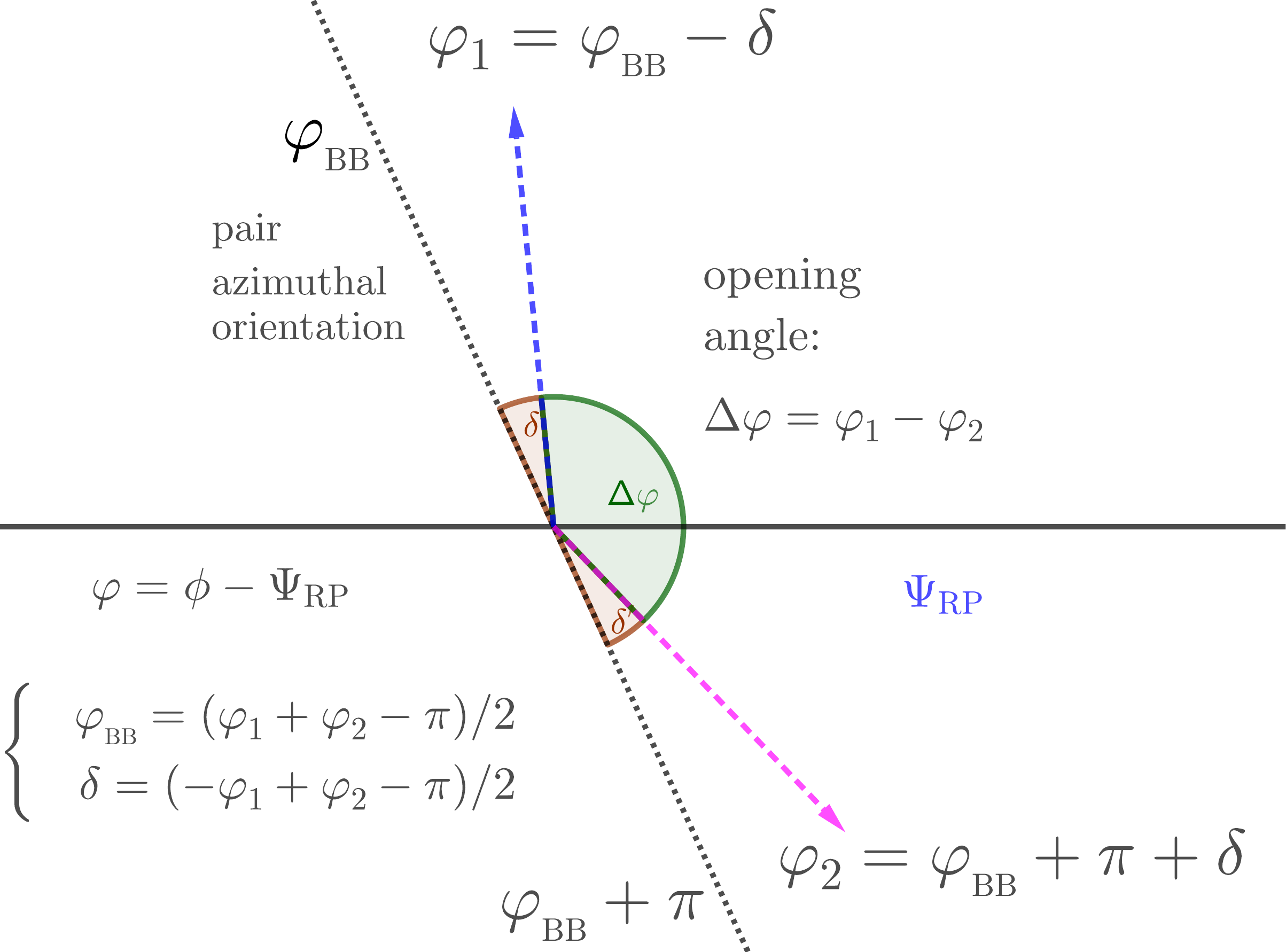}
	\caption{A sketch of ``back-to-back'' pair on the transverse plane.}
	\label{OpenAngle}
\end{figure}

We divide a heavy-ion collision event into three subevents according to the $\eta$ range, the east ($-1<\eta<-0.5$), middle ($-0.5\le\eta<0.5$), and west ($0.5\le\eta<1$) subevent.
The middle subevent is used to reconstruct the second-order event plane azimuthal angle ($\Psi_2$) 
as a proxy for that of the RP ($\Psi_{\text{RP}}$).
We form pairs of two charges, one from the west subevent and the other from the east subevent.
The middle subevent provides an $\eta$ gap between the pair of charges.
The opening angle between the two charges are required to be larger than a certain value (e.g.~$150^\circ$) 
to define as ``back-to-back'' pairs.
According to their charges, we classify those back-to-back pairs as either OS or SS pairs.
The azimuthal orientation of the back-to-back pairs is defined to be 
\begin{equation}
	\pbb = (\varphi_1 + \varphi_2 - \pi)/2
	,
\end{equation} 
where $\varphi_1$, $\varphi_2$ are the azimuthal angles of the two charges relative to $\Psi_{\text{RP}}$
(see Fig.~\ref{OpenAngle} for the various azimuthal angle definitions).
We count the numbers of the OS (SS) pairs, $\nos$ ($\nss$), as a function of $\pbb$.
We define our new observable as
\begin{equation}
	\rbb(\pbb) = \frac{\nos(\pbb) - \nss(\pbb)}{\nos(\pbb) + \nss(\pbb)}
	.
\end{equation}
If we expand this ratio by Fourier series, 
as we will show in Sec.~\ref{CMESignalExtraction},
the second-order coefficient of the Fourier expansion of this quantity
is a measure of the CME signal.


\subsection{CME signal extraction from $\rbb$} \label{CMESignalExtraction}

We first clarify analytically how $\rbb(\pbb)$ is sensitive to the CME signal.
The azimuthal distribution for the primordial pions can be written as
\begin{equation} \label{PrimordialPionAzimuthalDistribution}
	n^\pm(\varphi) \equiv \frac{\der N^\pm(\varphi)}{\der \varphi} = \frac{N^\pm}{2\pi}(1 \pm 2 a_1 \sin\varphi + 2 v_{2,\pi^\pm}\cos2\varphi)
	,
\end{equation}
where the superscript $\pm$ means the charge sign, 
and $N^\pm$ is the total number of primordial $\pi^\pm$ of the event.
The CME signal is described by the term $\pm 2 a_1\sin\varphi$.
A rough estimation is
$\langle a_1^2 \rangle \sim 10^{-4}$ in typical heavy ion collisions~\cite{Kharzeev:2007jp}.
Without loss of generality, we use $\varphi_1$ to denote a $\pi^+$ from the east subevent 
and $\varphi_2$ to denote a $\pi^-$ from the west subevent.
Transferring to pair variables $\pbb$ and $\delta$,
noting the Jacob determinant
$| \partial (\varphi_1, \varphi_2) / \partial (\pbb, \delta) | = 2,$
we obtain the pair distribution
\begin{equation}
\begin{split}
	&\nlp(\varphi_1) \nrn(\varphi_2) \der\varphi_1 \der\varphi_2 \\
	&= \nlp( \pbb - \delta) \nrn(\pbb + \pi + \delta) 2 \der\pbb \der\delta
	.
\end{split}
\end{equation}
Including the other case,
we have the OS pair density distribution
\begin{equation} \label{NosDist2}
\begin{split}
	&\nos(\pbb, \delta) \\
	=& 2\nlp( \pbb - \delta) \nrn(\pbb + \pi + \delta) \\
	&+ 2\nln( \pbb - \delta) \nrp(\pbb + \pi + \delta)\\
	&= \frac{\Nlp\Nrn + \Nln\Nrp}{2\pi^2}\Big[ 1 + 4a_1^2 \sin(\pbb+\delta)\sin(\pbb-\delta) \\
	&+ 4\vvp\vvn\cos2(\pbb+\delta)\cos2(\pbb-\delta)  \\
	&+ \cos2(\pbb+\delta) \\
	&\quad\times \left(\vvp + \vvn - 2a_1 (\vvp - \vvn)\sin(\pbb-\delta) \right) \\
	&+ \cos2(\pbb-\delta) \\
	&\quad\times \left(\vvp + \vvn + 2a_1 (\vvp - \vvn)\sin(\pbb+\delta) \right) \Big]
	.
\end{split}
\end{equation}
Assuming the event averages
\begin{equation}
	\langle \Nlp \Nrn \rangle = \langle \Nln\Nrp \rangle = \langle \Nlp\Nrp \rangle = \langle \Nln\Nrn \rangle = \langle N^2 \rangle
	,
\end{equation}
and intergrating over $\delta$ from $-\Delta$ to $\Delta$, we have
\begin{equation} \label{NosDist}
\begin{split}
	&\nos(\pbb) = \int_{\delta = -\Delta}^{\delta=\Delta} \nos(\pbb, \delta) \der\delta \\
	=& \frac{2 \langle N^2\rangle}{\pi^2} \Big[ \Delta + 2 \vvp\vvn \Delta \cos4\pbb + a_1^2 \sin2\Delta \\
	&+ \cos2\pbb ( -2a_1^2\Delta + (\vvp + \vvn)\sin2\Delta) \\
	&+ \frac{1}{2}\vvp\vvn\sin4\Delta \Big]
	.
\end{split}
\end{equation}
Similarly, we obtain the SS pair density distribution
\begin{equation}
\begin{split}
	&\nss(\pbb, \delta) \\
	=& 2\nlp( \pbb - \delta) \nrp(\pbb + \pi + \delta) \\
	&+ 2\nln( \pbb - \delta) \nrn(\pbb + \pi + \delta)
	,
\end{split}
\end{equation}
\begin{equation} \label{NssDist}
\begin{split}
	&\nss(\pbb) = \int_{\delta = -\Delta}^{\delta=\Delta} \nss(\pbb, \delta) \der\delta \\
	=& \frac{2\langle N^2\rangle}{\pi^2} \Big[ \Delta + (\vvp^2 + \vvn^2)\Delta\cos4\pbb - a_1^2 \sin2\Delta \\
	&+ \cos2\pbb (2 a_1^2 \Delta + (\vvp+\vvn)\sin2\Delta)\\
	& + \frac{1}{4}(\vvp^2 + \vvn^2)\sin4\Delta \Big]
	.
\end{split}
\end{equation}
The difference and sum are, respectively,
\begin{equation}
\begin{split}
	&\nos(\pbb) - \nss(\pbb) \\
	=& \frac{2\langle N^2\rangle}{\pi^2} \Big[ -4a_1^2 \Delta \cos2\pbb + 2a_1^2\sin2\Delta \\
	&- (\vvp-\vvn)^2\Delta \cos4\pbb \\
	&- \frac{1}{4}(\vvp-\vvn)^2\sin4\Delta \Big]
	,
\end{split}
\end{equation}
\begin{equation}
\begin{split}
	&\nos(\pbb) + \nss(\pbb) \\
	=& \frac{2\langle N^2\rangle}{\pi^2} \Big[ 2\Delta + (\vvp+\vvn)^2 \Delta \cos4\pbb \\
	&+ (\vvp+\vvn)\cos2\pbb\sin2\Delta \\ 
	&+ \frac{1}{4}(\vvp+\vvn)^2\sin4\Delta \Big]
	.
\end{split}
\end{equation}
Our new observable is the ratio and we expand it into Fourier series
\begin{equation} \label{BackToBackRatio}
	\rbb(\pbb) = \frac{\nos(\pbb) - \nss(\pbb)}{\nos(\pbb) + \nss(\pbb)}
	= \sum_{k=0}^{+\infty} c_k \cos(k\pbb)
	.
\end{equation}
Noticing that $(\vvp+\vvn)$ is small ($\sim0.1$), up to the first order of $(\vvp+\vvn)$,
the coefficient of $\cos2\pbb$ is
\begin{equation} \label{C2}
\begin{split}
	&c_2 \approx a_1^2 \left( -2 - (\vvp+\vvn)\frac{\sin^2 2\Delta}{\Delta^2} \right) \\
	&+(\vvp+\vvn) (\vvp-\vvn)^2 \frac{(2\Delta+\sin4\Delta)\sin2\Delta}{8\Delta^2}
	.
\end{split}
\end{equation}
If we require the opening angle to be larger than $150^\circ$ for the back-to-back pairs, then $\Delta=15^\circ$,
\begin{equation} \label{NumberC2}
\begin{split}
	c_2 \approx& a_1^2 \left( -2 -3.648(\vvp+\vvn) \right) \\
	&+ 1.267(\vvp + \vvn)(\vvp-\vvn)^2
	.
\end{split}
\end{equation}
The second term is not related to the CME;
taking 
$|\vvp-\vvn| \sim 10^{-3}, \quad (\vvp+\vvn) \sim 10^{-1}$,
its magnitude is on the order of $10^{-7}$.
For a CME signal of $a_1 \ge 10^{-3}$, 
$a_1^2$ dominates over the primordial flow effects in $c_{2}$,
indicating that $c_{2}$ is a good measure of the CME.

Similarly, the coefficient of the constant term ($k=0$) is
\begin{equation}
\begin{split}
	c_0 =& \frac{\sin2\Delta}{4\Delta} \big( 4 a_1^2 (1 + \vvp + \vvn) \\
	&- (\vvp - \vvn)^2 \cos2\Delta \big)
	,
\end{split}
\end{equation}
and for $\Delta=15^\circ$,
\begin{equation} \label{NumberC0}
\begin{split}
	c_0 \approx 1.910 a_1^2 (1 + \vvp + \vvn) - 1.654 (\vvp - \vvn)^2
	.
\end{split}
\end{equation}
Note that $c_{2}$ and $c_{0}$ are both sensitive to the CME,
with similar sensitivities.
It will be shown later, however, that $c_{0}$ is also sensitive to the backgrounds.
Those backgrounds are mainly from the low $p_{T}$ resonance decays whose decay daughters are back-to-back.
The $c_{2}$ is less sensitive to those backgrounds because their $v_2$ at low $p_{T}$ is small.


\begin{figure*}
	\includegraphics[width=0.4\linewidth]{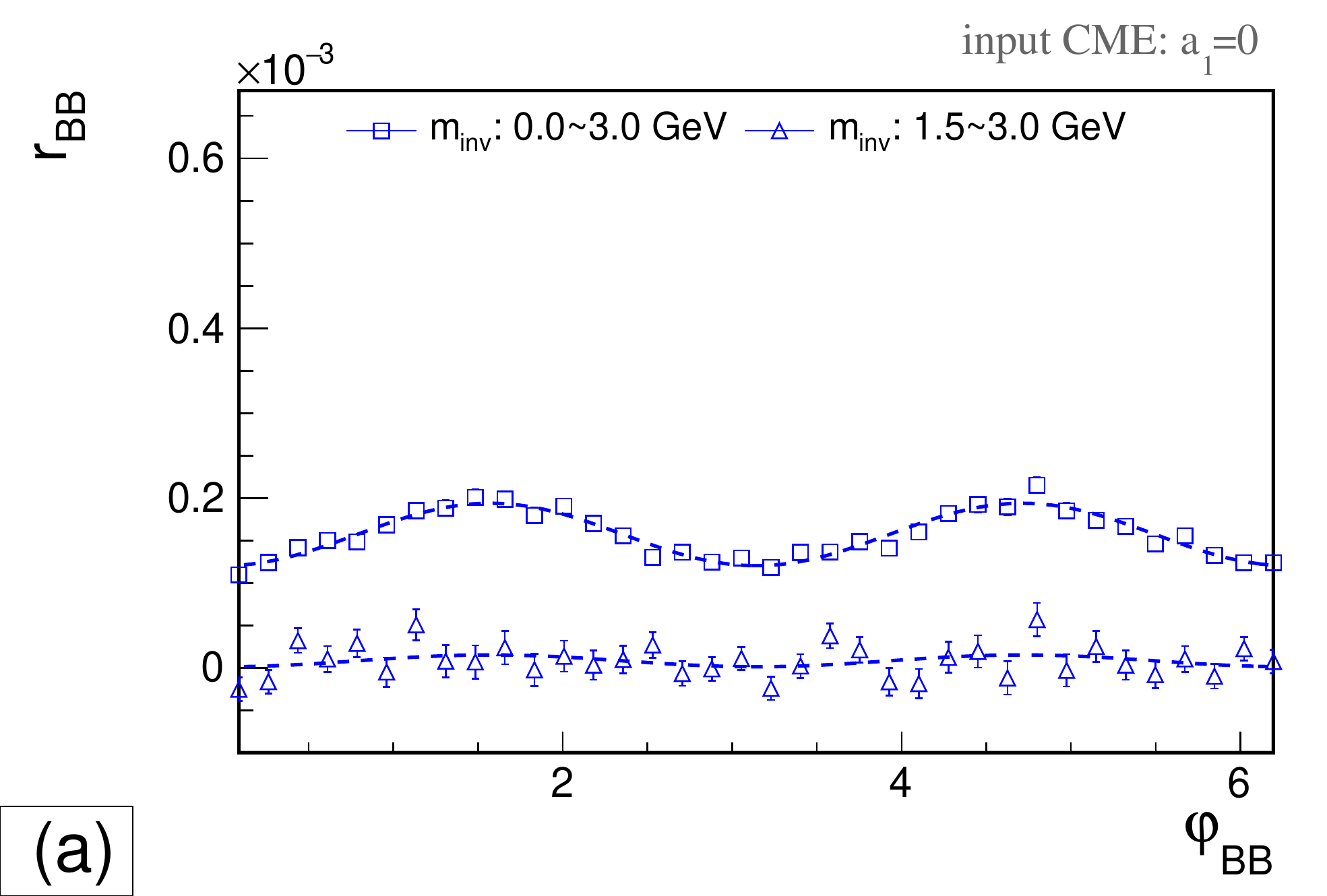} \hspace{0.05\linewidth}
	\includegraphics[width=0.4\linewidth]{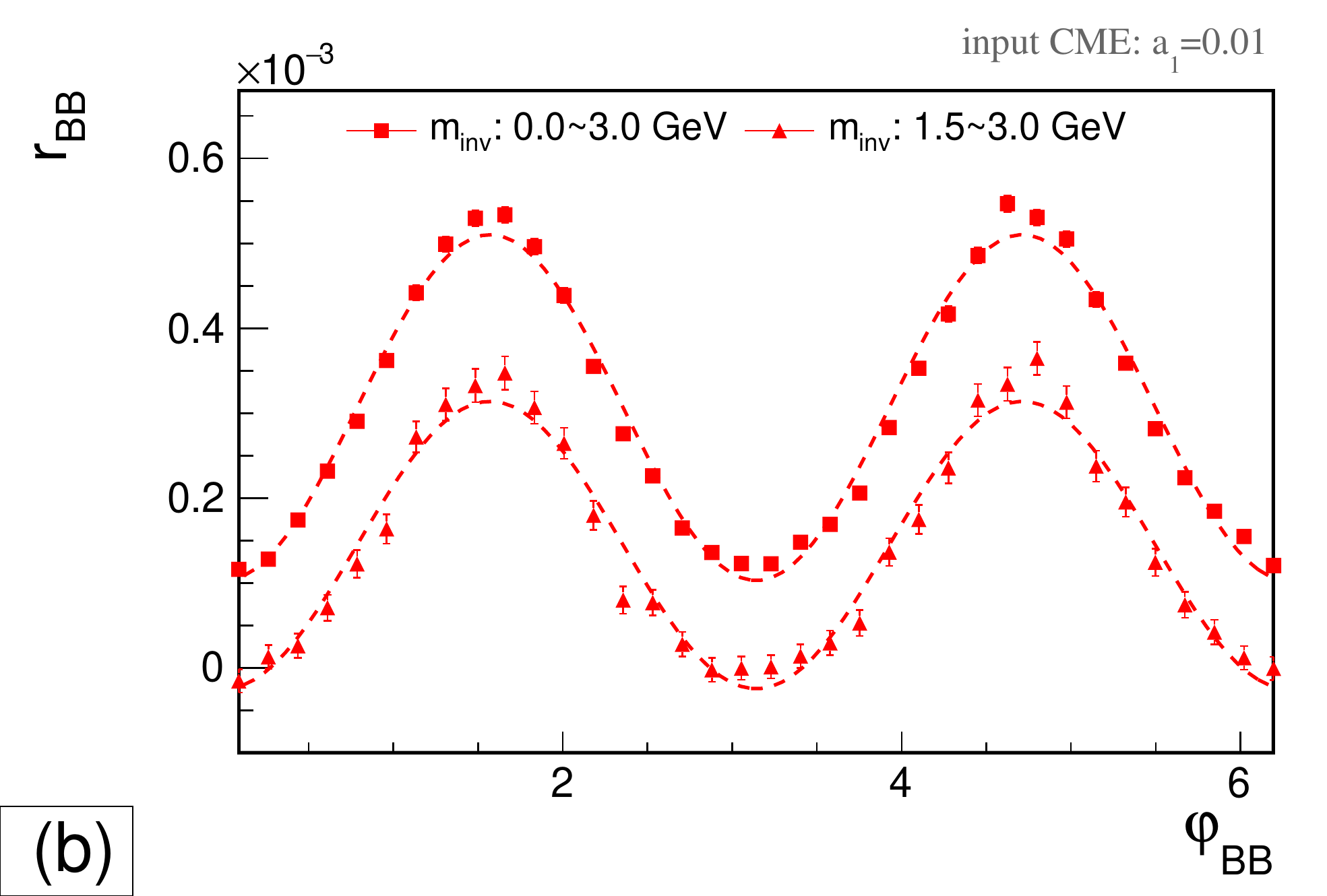}
	\caption{The distribution of the back-to-back relative-excess observable $\rbb(\pbb)$ in the toy model simulations with input CME (a) $a_1=0$ and (b) $a_1=0.01$.}
	\label{ToyModelObsDist}
	\includegraphics[width=0.4\linewidth]{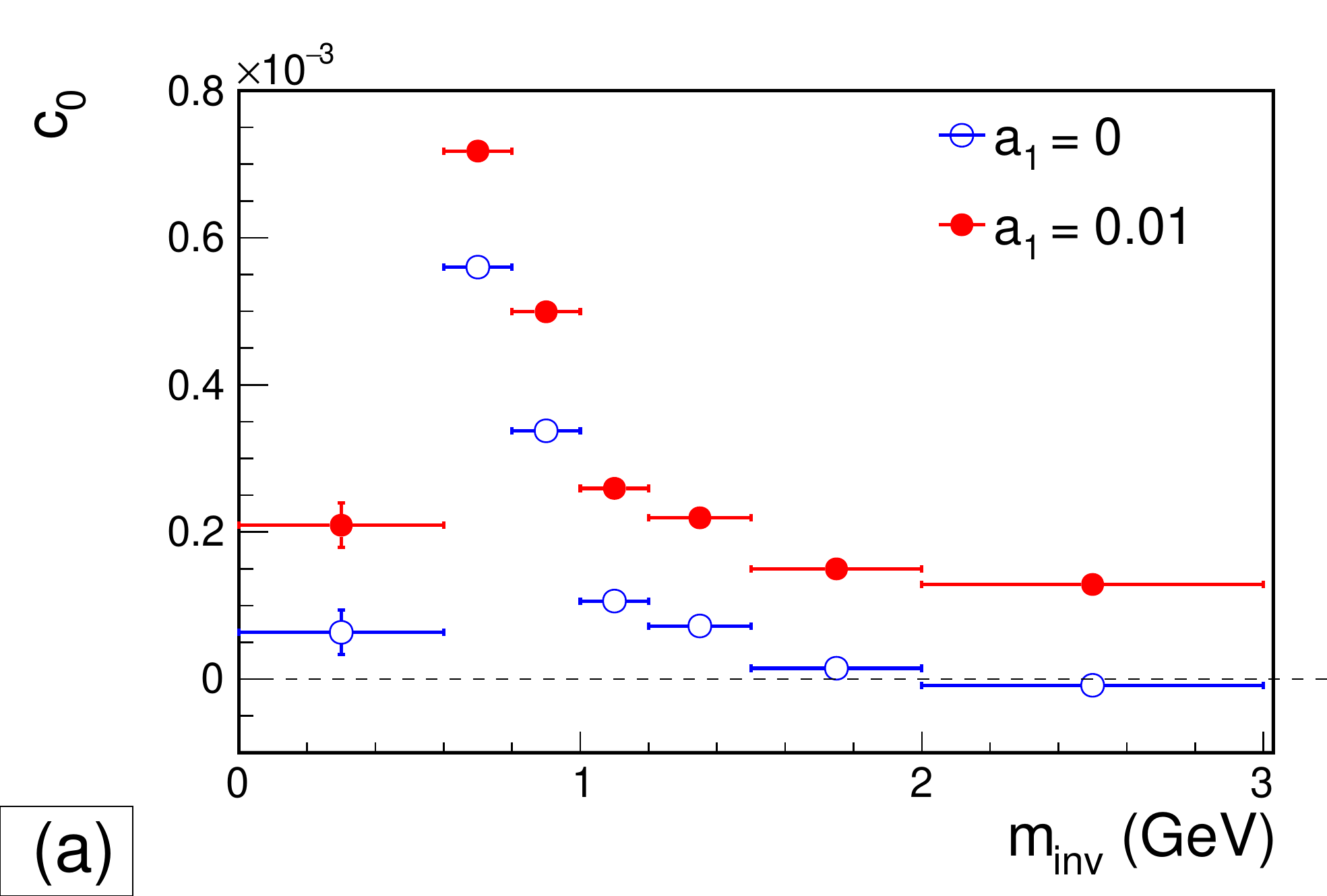} \hspace{0.05\linewidth}
	\includegraphics[width=0.4\linewidth]{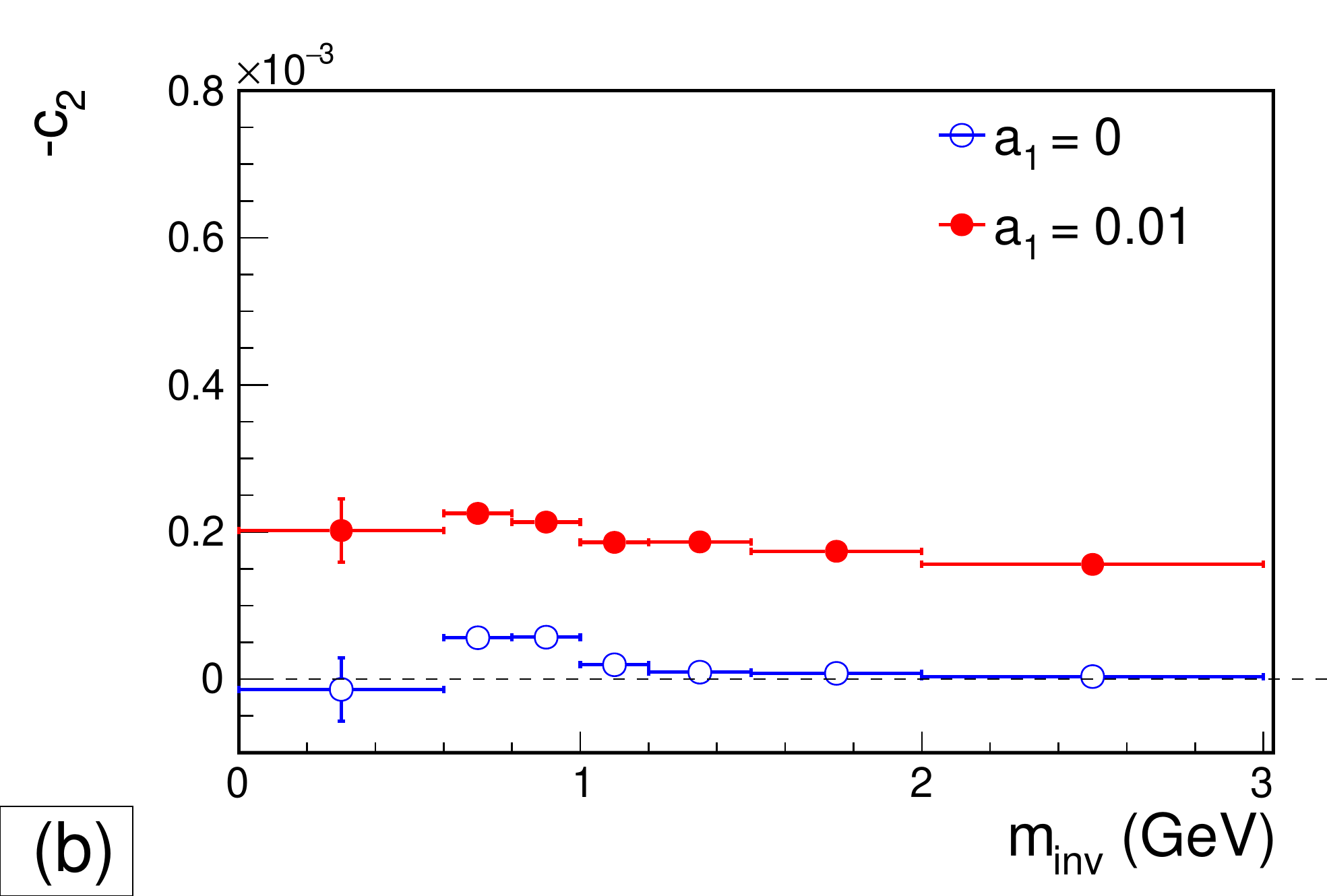}
	\caption{The fitted Fourier coefficients, (a) $c_0$ and (b) $-c_2$, to the back-to-back relative-excess observable $\rbb(\pbb)$ in the toy model with various CME inputs.}
	\label{ToyModelFitParC}
\end{figure*}

\subsection{Comparison to the back-to-back $\gdf$ observable, $\gdfbb$}

The $\Delta\gamma$ observable is frequently used in heavy-ion collisions to search for the CME,
\begin{equation} \label{GammaRP}
\begin{split}
	\gdf =& \gos - \gss,\\
	\gamma =& \langle \cos(\varphi_1 + \varphi_2) \rangle
	.
\end{split}
\end{equation}
To see the relationship between $\rbb$ and $\gdf$,
we will apply the same ``back-to-back'' requirement to the pairs in $\gdf$, 
denoted as $\gdfbb$.
For back-to-back pairs,
$\cos(\varphi_1 + \varphi_2) = -\cos(2\pbb)$.
The correlators $\gos$ and $\gss$ can be simplified into
\begin{equation}
\begin{split}
	\gos =& - \frac{ \int \cos(2\pbb) \nos(\pbb) \der\pbb }{ \int \nos(\pbb) \der\pbb }\\
	=& \frac{2 a_1^2 \Delta - (\vvp+\vvn) \sin2\Delta}{2\Delta + 2a_1^2\sin2\Delta + \vvn\vvp\sin4\Delta}, \\
	\gss =& - \frac{ \int \cos(2\pbb) \nss(\pbb) \der\pbb }{ \int \nss(\pbb) \der\pbb } \\
	=& \frac{ -2 a_1^2 \Delta - (\vvp+\vvn) \sin2\Delta}{ 2\Delta - 2a_1^2\sin2\Delta + \frac{1}{2}(\vvn^2+\vvp^2)\sin4\Delta}
	.
\end{split}
\end{equation}
The difference to the first order of $(\vvp+\vvn)$ is therefore
\begin{equation} \label{GammaB2B}
\begin{split}
	\gdfbb =& \gos - \gss \\
	\approx& a_1^2 \left( 2 + (\vvp+\vvn)\frac{\sin^2 2\Delta}{\Delta^2}  \right) \\
	&-(\vvp+\vvn)(\vvp-\vvn)^2\frac{\sin2\Delta\sin4\Delta}{8\Delta^2}
	.
\end{split}
\end{equation}
With $\Delta=15^\circ$, it becomes
\begin{equation} \label{NumberGammaB2B}
\begin{split}
	\gdfbb =& \gos - \gss \\
	\approx& a_1^2 \left( 2 + 3.648(\vvp+\vvn) \right) \\
	&-0.790(\vvp+\vvn)(\vvp-\vvn)^2
	.
\end{split}
\end{equation}
Comparing Eqs.~\ref{GammaB2B} and \ref{NumberGammaB2B} to Eqs.~\ref{C2} and \ref{NumberC2}, 
it is clear that $\gdfbb$ and $\rbb$ have similar sensitivity to the CME.
The $\rbb$ observable is directly related to $\gdfbb$.
Only the back-to-back pairs are used in these two observables,
so the backgrounds among the close pairs are reduced.


\section{Results} \label{Results}

In this section, we show 
the back-to-back $\rbb(\pbb)$ and back-to-back $\gdfbb$ observables calculated 
from a toy model (with/without input CME) simulations.


\subsection{Toy-model simulation}

We use a toy model including the primordial pions and the $\rho$ meson decays to study 
the sensitivities of $\rbb$ to CME signal and resonance backgrounds.
This toy model has been used for CME background studies in Ref.~\cite{Wang:2016iov, Feng:2018so}.
Both the resonance decays and primordial pions have the $p_T$ distributions and $v_2(p_T)$ obtained from Au+Au measurements corresponding to centrality $40\%\sim50\%$~\cite{Adams:2003cc, Adler:2003qi, Adams:2003xp, Abelev:2008ab, Adams:2004bi, Adare:2010sp, Dong:2004ve, Adamczyk:2015lme, Agashe:2014kda, Abelev:2009gu, Wang:2016iov}.

To simulate the CME signal in the toy model, we input the coefficient $a_1$ when generating the primordial pions from the azimuthal distribution (Eq.~\ref{PrimordialPionAzimuthalDistribution}).
Two cases are studied, one without CME input ($a_1=0$), and the other with $1\%$ CME input ($a_1=0.01$). 
Each case has $2\times10^{9}$ events.
The tracks are selected with transverse momentum $0.2\text{ GeV} < p_T < 2.0\text{ GeV}$ and pseudorapidity $-1.0 < \eta < 1.0$.
Figure~\ref{ToyModelObsDist} shows the $\rbb(\pbb)$ distributions for the two cases.
The case with finite CME shows larger amplitude and modulation than the case without,
indicating the sensitivity of the $\rbb(\pbb)$ observable to the CME.
The case without CME shows some finite amplitude and modulation, at low $m_{\text{inv}}$, 
indicating that the observable still has some background contamination.
In order to further suppress resonance backgrounds, we also show the $\rbb$ distributions 
with the invariant mass range $1.5 \text{ GeV} < m_\text{inv} < 3.0 \text{ GeV}$.
The result is consistent with zero as expected.

\begin{table*}[] 
\begin{tabular}{|c|c|c|c|c|}
\hline
$m_{\text{inv}}$ range (GeV) & input $a_1$ & \multicolumn{2}{|c|}{Fourier coefficients ($\times10^{-4}$)} & extracted $a_1$ ($\times10^{-2}$) \\ \hline
\multirow{4}{*}{$0.0 \sim 3.0$} & \multirow{2}{*}{$0$} & $\ \  c_0$ & $1.57 \pm 0.01$ & $0.87\pm0.01$ \\ \cline{3-5}
 & & $-c_2$ & $0.37 \pm 0.02$ & $0.39 \pm 0.02$ \\ 
 \cline{2-5}
 & \multirow{2}{*}{$0.01$} & $\ \ c_0$ & $3.07 \pm 0.01$ & $1.21\pm0.01$ \\ \cline{3-5}
 & & $-c_2$ & $2.04 \pm 0.02$ & $0.93 \pm 0.01$ \\ \hline
 \multirow{4}{*}{$1.5 \sim 3.0$} & \multirow{2}{*}{$0$} & $\ \ c_0$ & $0.08 \pm 0.03$ & $0.20\pm0.07$ \\ \cline{3-5}
 & & $-c_2$ & $0.07\pm0.04$ & $0.17\pm0.10$ \\ 
 \cline{2-5}
 & \multirow{2}{*}{$0.01$} & $\ \ c_0$ & $1.45 \pm 0.03$ & $0.83 \pm 0.02$ \\ \cline{3-5}
 & & $-c_2$ & $1.69 \pm 0.04$ & $0.85\pm0.02$ \\ \hline
\end{tabular}
\caption{The fitted Fourier coefficients $c_0$ and $-c_2$ are shown for the $\rbb(\pbb)$ distributions from the toy-model simulations with/without CME signal input. Two invariant mass ranges are shown. 
If we set $(\vvp+\vvn) \approx 0.1$, ignore the $(\vvp-\vvn)^2$ terms in Eqs.~\ref{NumberC2}~and~\ref{NumberC0}, 
and assume zero resonance background in the toy-model simulations, 
the $a_1$ can be extracted from $c_0$ and $c_2$, respectively.
These extracted $a_{1}$ values are also listed,
to be compared to the input $a_{1}$.}
\label{FittingParTable}
\end{table*}

We fit the $\rbb(\pbb)$ distributions to Eq.~\ref{BackToBackRatio}.
Figure~\ref{ToyModelFitParC} shows the fitted Fourier coefficients $c_0$ and $-c_2$, respectively,
as a function of $m_{\text{inv}}$.
The $c_0$ has strong sensitivity to both signal and background.
Although still affected by the residual resonance backgrounds,
the $-c_2$ has better sensitivity to CME than $c_0$ and less sensitivity to background.
To illustrate our results more quantitatively,
we list the fitted coefficients $c_0$ and $-c_2$ in Table~\ref{FittingParTable}.
Also listed are the $a_1$ values extracted from $c_{0}$ and $-c_{2}$,
via Eqs.~\ref{NumberC0} and \ref{NumberC2}, respectively,
ignoring the presence of backgrounds.
Due to resonance backgrounds in the low $m_{\text{inv}}$ range,
the extracted $a_1$ are large with $0.0 \text{ GeV} < m_{\text{inv}} < 3.0 \text{ GeV}$,
no matter whether the input $a_1$ are zero or not.
In the range $1.5 \text{ GeV} < m_{\text{inv}} < 3.0 \text{ GeV}$,
with zero input $a_1$,
the extracted $a_1$ values are also close to zero;
the small deviations from zero are due to residual resonance backgrounds.
With input $a_1=0.01$,
the extracted $a_1$ values in the high $m_{\text{inv}}$ range are nonzero, close to the input;
again, the differences are due to residual resonance backgrounds.
However, under this condition, the extracted $a_1$ values are smaller than the inputs.
This is because there are pairs composed of pions from uncorrelated sources
(one primordial pion and one resonance pion, or two pions from two different resonance decays),
whose zero contributions are averaged in $c_0$, $-c_2$. 
The dilution from those uncorrelated pairs reduces the extracted $a_1$ values.


\subsection{Comparison among $\gdf$, $\gdfbb$, and $-c_2$}

\begin{figure}
	\includegraphics[width=0.85\linewidth]{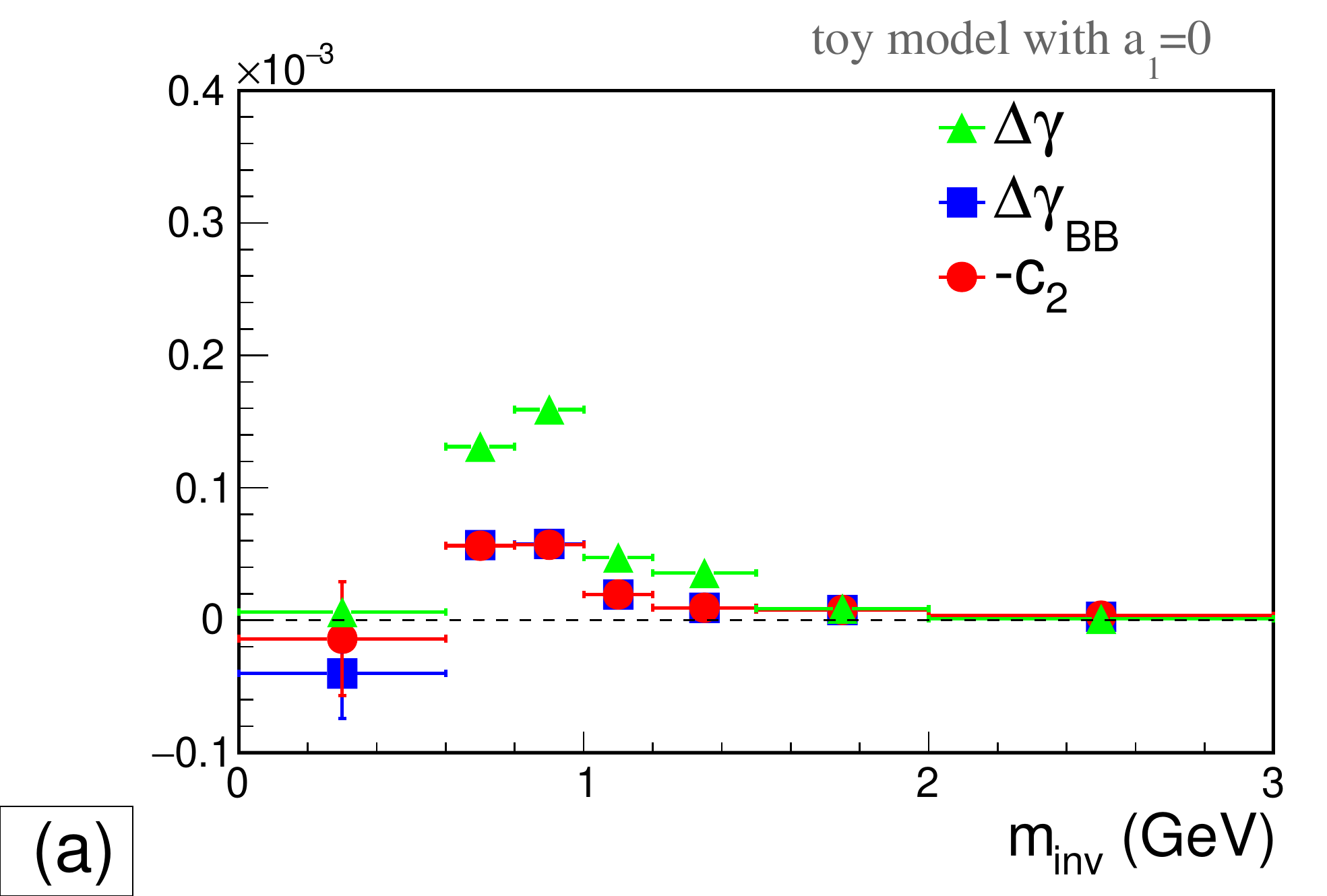}
	\includegraphics[width=0.85\linewidth]{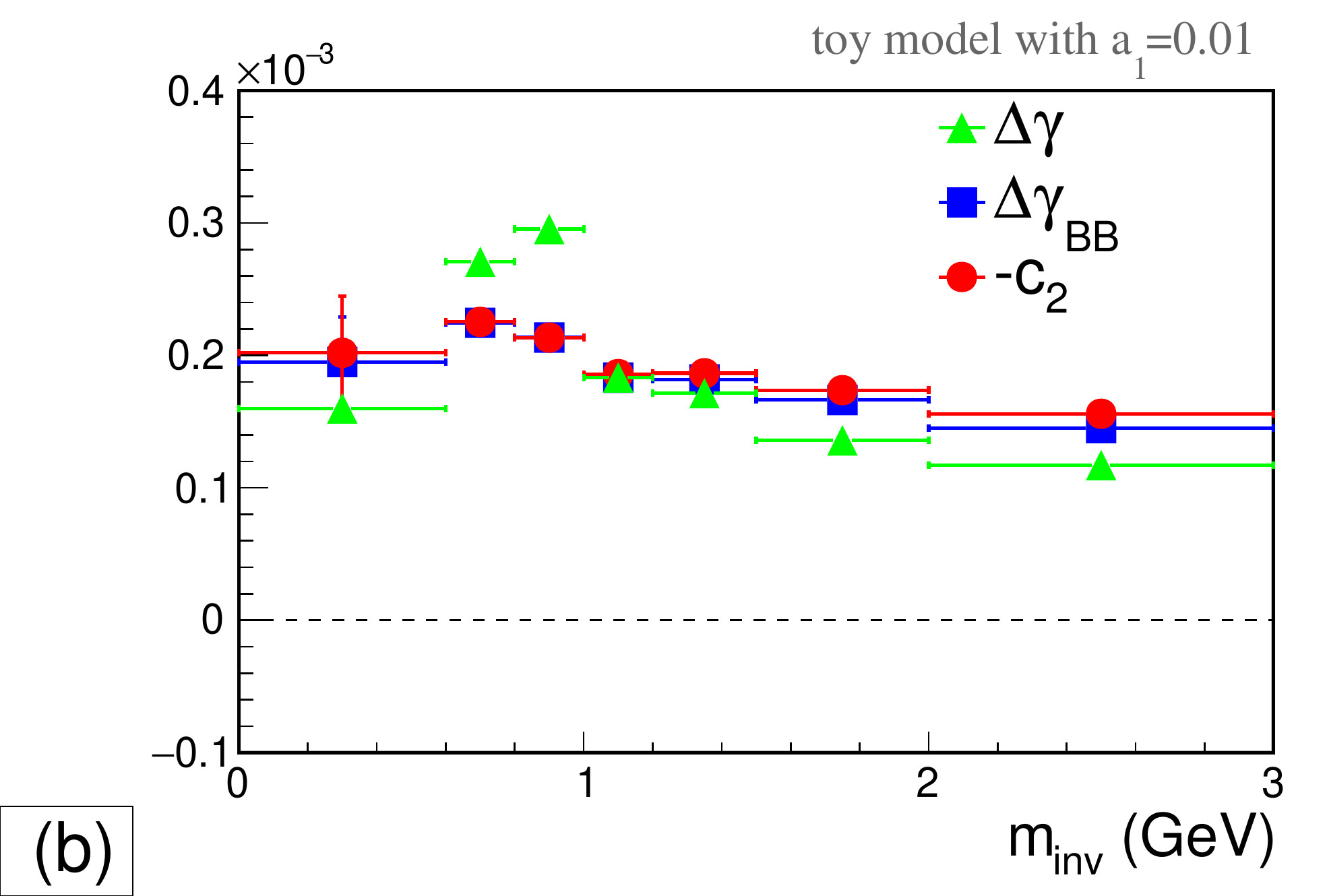}
	\caption{Comparison among the inclusive $\gdf$, the back-to-back $\gdfbb$, and the Fourier coefficient $-c_2$ of the $\rbb(\pbb)$ distribution in different simulations.}
	\label{CompGammaC2}
\end{figure}

We also calculate the inclusive $\gdf$ and back-to-back $\gdfbb$ observables in our model studies.
Figure~\ref{CompGammaC2} compares the results of those three observables.
It is found that $\gdfbb$ and $-c_2$ are very close to each other.
This indicates that the $\rbb$ and $\gdfbb$ observables are nearly the same,
as expected from Eqs.~\ref{NumberC2} and \ref{NumberGammaB2B}.
With zero CME input ($a_1=0$) in the toy model simulation (Fig.~\ref{CompGammaC2}a),
the inclusive $\gdf$ is further away from zero than the other two observables
in the invariant mass range $0.6 \sim 1.5 \text{ GeV}$ where resonance contributions are large.
This shows that the inclusive $\gdf$ is more significantly affected by the resonance backgrounds.
In the high mass region where resonance contributions are small,
all three observables approach to zero as expected.
With nonzero CME input ($a_1=0.01$) in the toy model simulation (Fig.~\ref{CompGammaC2}b), 
the three observables are all away from zero.
The inclusive $\gdf$ is lower than the other two in the mass range $1.5 \text{ GeV} \sim 3.0 \text{ GeV}$
where there is not much resonance contribution.
This is because the back-to-back CME signal is diluted more in the inclusive $\gdf$ by including close pairs from backgrounds.
This can also be explained by the analytical calculations in Eq.~\ref{GammaB2B} 
by assigning $\Delta=90^\circ$ for the inclusive $\gdf$ and $\Delta=15^\circ$ for $\gdfbb$.

\section{Summary} \label{Summary}

In this paper, we propose a new observable to search for the CME,
called the back-to-back relative-excess observable of OS to SS pairs ($\rbb$),
as a function of the pair azimuthal orientation ($\pbb$).
The charge pairs used in this observable are required to be back-to-back: 
opening angle larger than $150^\circ$ on the transverse plane; 
they are taken from different $\eta$ ranges with a $\Delta\eta$ gap to further reduce backgrounds.
As a result, the backgrounds (such as resonance decays) contributing mostly to the close pairs can be reduced.
A modulation of the form $\cos2\pbb$ in the observable can indicate a CME signal,
which is described by the second-order coefficient $c_2$ in Fourier expansion.

We use a toy model simulation without input CME ($a_1=0$) and with $1\%$ input CME ($a=0.01$), 
and calculate the observable from the simulated data.
The coefficient $c_2$ is close to zero when there is no input CME,
whereas it is far from zero with $1\%$ input CME.

To relate the new observable to the previous $\Delta\gamma$ observable,
we apply the same back-to-back pair requirement to the definition of $\Delta\gamma$ to obtain $\gdfbb$.
We use analytical calculations and toy-model simulations to show that $\gdfbb$ is nearly identical to $-c_2$.
Both are more sensitive to the CME and less sensitive to resonance backgrounds than the inclusive $\gdf$ observable.


\section*{Acknowledgments}

This work is supported in part by the U.S.~Department of Energy Grant No.~DE-SC0012910 
and the National Natural Science Foundation of China Grant No.~11847315.


\bibliography{./ref}


\end{document}